\documentclass[11pt]{article}
\pdfoutput=1
\usepackage{amsmath,amssymb,amsfonts,amsthm,bm,bbm,cancel,wasysym}
\usepackage{epsfig,graphics,graphicx,epstopdf,caption,subcaption}
\usepackage[numbers,sort&compress]{natbib}
\graphicspath{{Charts/}}
\usepackage{array,booktabs,colortbl,colordvi,multirow}
\usepackage{colordvi,color,xcolor}
\usepackage{hyperref}
\usepackage{rotating}
\usepackage{comment}
\usepackage{feynmf}

\usepackage[symbol]{footmisc}
\renewcommand{\thefootnote}{\fnsymbol{footnote}}

\begin{document}

\begin{center}

{\Large {\bf Lyman-$\alpha$ limit on axion-like cold dark matter}}\\
%\vspace*{0.15cm}

\vspace*{0.75cm}

{Zixuan Xu and Sibo Zheng}

\vspace{0.5cm}
{Department of Physics, Chongqing University, Chongqing 401331, China}
\end{center}
\vspace{.5cm}

%--------------------------------------------- ABSTRACT ---------------------------------------------%
\begin{abstract}
\noindent
Using low redshift data on astrophysical reionization, 
we report new Lyman-$\alpha$ limit on axion-like particle (ALP) as cold dark matter in ALP mass range of $m_{a}\sim 30-1000$ eV.
Compared to the Leo T and soft-X ray bound, this limit is so far the most stringent in the ALP mass range of $m_{a}\sim 375-425$ eV and complementary in the  ALP mass range otherwise. 
Combing these limits, we show new exclusion limits on $m_a$ for the ALP DM from either misalignment or freeze-in mechanism.
\end{abstract}

\renewcommand{\thefootnote}{\arabic{footnote}}
\setcounter{footnote}{0}
\thispagestyle{empty}
\vfill
\newpage
\setcounter{page}{1}

\tableofcontents

%%%%%%%%%%%%%%%%%%%%%%%%%%%%
\section{Introduction}
Dark matter (DM) has shown its existence through gravitational interactions.
To uncover the particle nature of DM, various experiments aiming to detect both weakly interacting massive particle (WIMP)
and non-WIMP have been put in place. 
For the later there are several examples including millicharged particle \cite{Holdom:1985ag,Feldman:2007wj,Yin:2023jql},
sterile neutrino \cite{Asaka:2005cn,Becker:2018rve,Chianese:2018dsz,Datta:2021elq},  scalar \cite{McDonald:2001vt}, and QCD axion \cite{Peccei:1977hh, Peccei:1977ur,Weinberg:1977ma,Wilczek:1977pj} to solve the strong CP problem.
A common feature of these hypothetical particles is a feeble coupling to the Standard Model (SM),
which allows them to hide in parameter regions beyond conventional experimental reaches.

Astrophysical and cosmic rays are few of tools being applicable to test non WIMP-like DM.
For DM mass larger than $\sim$ keV scale, $X/\gamma$-rays \cite{Cadamuro:2011fd,Vertongen:2011mu,Essig:2013goa,Cohen:2016uyg} due to DM decays inside specified galaxies cannot excess the known backgrounds.
For DM mass above $\sim$ eV scale, observations of cosmic rays including Lyman-$\alpha$ forest and 21-cm signal provide data about temperature and reionization of the intergalactic medium (IGM), allowing us to place constraints on DM annihilation or decay into photons or electron-positron pair.
So far,  cosmic rays have been mainly used to constrain DM with mass above keV scale. 
Refs.\cite{Liu:2016cnk,Liu:2020wqz,Xu:2024uas} show that current Lyman-$\alpha$ data has already excluded DM lifetime (cross section of annihilation into electron-positron) smaller (larger) than $\sim 10^{25}-10^{26}$ sec ($10^{-30}-10^{-28}$ cm$^{3}$s$^{-1}$).
Refs.\cite{Xu:2024uas,Sun:2023acy,Facchinetti:2023slb,Mitridate:2018iag}  show that future sensitivities of 21-cm signal can provide stronger constraints than the Lyman-$\alpha$ forest.

In practice, the cosmic rays are more useful to probe DM with mass below keV scale such as axion-like particle (ALP) - an extension of QCD axion.
See e.g. \cite{Marsh:2015xka,Adams:2022pbo} for reviews.  
In this work, we report new Lyman-$\alpha$ limit on the ALP DM within ALP mass range of eV - keV.
Starting with the ALP interactive Lagrangian
\begin{eqnarray}\label{Lag}
\mathcal{L}_{\rm{eff}}=-\frac{g_{a\gamma\gamma}}{4} a F_{\mu\nu}\tilde{F}^{\mu\nu}+\cdots,
\end{eqnarray}
where $a$ is the ALP and $F$ is the electromagnetic field strength respectively,
one obtains the ALP lifetime
\begin{eqnarray}
\label{lifetime}
\tau_{a} \approx 8.5\times 10^{35} \rm{sec} \left(\frac{g_{a\gamma\gamma}}{10^{-10}~\rm{GeV}}\right)^{-2}\left(\frac{m_{a}}{250~\mu\rm{eV}}\right)^{-3},
\end{eqnarray}
where interactions neglected in eq.(\ref{Lag}) do not affect the value of $\tau_a$ in the ALP mass range of $m_{a}< 2m_{e}$ considered here. 
Eq.(\ref{lifetime}) points to $\tau_{a}$ ranging from $10^{25}$ sec to $10^{30}$ sec in parameter regions
with $g_{a\gamma\gamma}\sim 10^{-12}-10^{-11}$ GeV$^{-1}$ and $m_{a}\sim 1-10^{3}$ eV,
 which are not yet excluded by current experiments limits.
We will show that these parameter regions are in the reach of current Lyman-$\alpha$ data, 
and uncover the physical implications to the ALP DM from either misalignment \cite{Abbott:1982af, Preskill:1982cy,Dine:1982ah,Turner:1983he} or freeze-in mechanisms \cite{Hall:2009bx}.

\section{Lyman-$\alpha$ limit}

$\underline{\mathbf{Stimulation}}$. In the late-time Universe the evolution of IGM ionization fraction and temperature is described by \cite{Liu:2019bbm} 
\begin{eqnarray}
\frac{dx_{\rm{HII}}}{dz}&=&\frac{dt}{dz}\left(\Lambda_{\rm{ion}}-\Lambda_{\rm{rec}}+\Lambda^{\rm{DM}}_{\rm{ion}}\right),\label{DE1}\\
\frac{dT_{k}}{dz}&=&\frac{2}{3}\frac{T_{k}}{n_{b}}\frac{dn_{b}}{dz}-\frac{T_{k}}{1+x_{e}}\frac{dx_{e}}{dz}+\frac{2}{3k_{B}(1+x_{e})}\frac{dt}{dz}\left(\sum_{p}\epsilon^{p}_{\rm{heat}}+\epsilon_{\rm{heat}}^{\rm{DM}}\right),\label{DE2}
\end{eqnarray}
respectively, where $x_{\rm{HII}}=n_{H^{+}}/n_{H}$ is the ionization fraction with $n_{H}$ ($n_{H^{+}}$) the number density of (ionized) hydrogen, $T_{k}$ the matter (baryon) temperature, $n_b$ the baryon number density, $k_B$ the Boltzmann constant, $z$ the redshift, 
$\Lambda_{\rm{ion}}$ the ionization rate, $\Lambda_{\rm{rec}}$ the recombination rate, $\epsilon^{p}_{\rm{heat}}$  the Compton scattering and astrophysical source induced heating rate.

The ALP DM decay induced terms in eqs.(\ref{DE1}) and (\ref{DE2}) are given by \cite{Liu:2019bbm}
\begin{eqnarray}
\Lambda^{\rm{DM}}_{\rm{ion}}&=& \mathcal{F}_{\rm{H}}\frac{\epsilon^{\rm{DM}}_{\rm{HII}}}{E^{\rm{HI}}_{\rm{th}}}+\mathcal{F}_{\rm{He}}\frac{\epsilon^{\rm{DM}}_{\rm{HeII}}}{E^{\rm{HeI}}_{\rm{th}}},\label{DMrate1}\\
\epsilon^{\rm{DM}}_{c}&=&f_{c}(x_{e},z)\frac{1}{n_{b}}\left(\frac{dE(z)}{dtdV}\right)_{\rm{inj}},\label{DMrate2}
\end{eqnarray}
where $f_{c}(x_{e},z)$ are the deposition fractions, with deposition channels including IGM heating (c=heat), hydrogen ionization (c = HII), helium single or double ionization (c = HeII or HeIII), and neutral atom excitation (c = exc), 
$\mathcal{F}_{j}$ refers to the number fraction of each species $j$, $E^{j}_{\rm{th}}$ is the energy for ionization,
and the DM induced energy injection rate is 
\begin{eqnarray}\label{DMrate}
\left(\frac{dE(z)}{dtdV}\right)_{\rm{inj}}=
\frac{\rho_{\rm{DM},0}(1+z)^{3}}{\tau_{\rm{a}}},
\end{eqnarray}
with $\rho_{\rm{DM},0}$ the observed DM relic density.

Using the publicly available package \texttt{DarkHistory} \cite{Liu:2019bbm}, 
we calculate the DM decay induced effects on the evolution of IGM parameters with $m_{a}$ and $\tau_{a}$ as free input parameters.

$\underline{\mathbf{Data}}$. Reionization is believed to occur due to the astrophysical source.
Therefore the data \cite{Walther:2018pnn,Gaikwad:2020art} about $T_{k}$ in the low redshift range of $z\sim 4-7$,
together with the Planck data \cite{Planck:2018vyg} about $x_e$, 
allows us to fix the astrophysical contribution in eqs.(\ref{DE1}) and (\ref{DE2}). 
To do so, we follow \cite{Liu:2020wqz} to adopt the FlexKnot model to parametrize the astrophysical source contribution to $\Lambda_{\rm{ion}}$ in eq.(\ref{DE1}) and photoheated prescription to parametrize the astrophysical source contribution to $\epsilon^{p}_{\rm{heat}}$ in eq.(\ref{DE2}), respectively.
Given the astrophysical reionization established, 
it is straightforward to carry out the Lyman-$\alpha$ limit on the DM induced effects on the IGM parameters. 

%%%%%%%%%%%%%%%%%%%%%%%%%%%%%%%%%%%%%%%%%%%%%%%%
\begin{figure}[htb!]
\centering
\includegraphics[width=15cm,height=15cm]{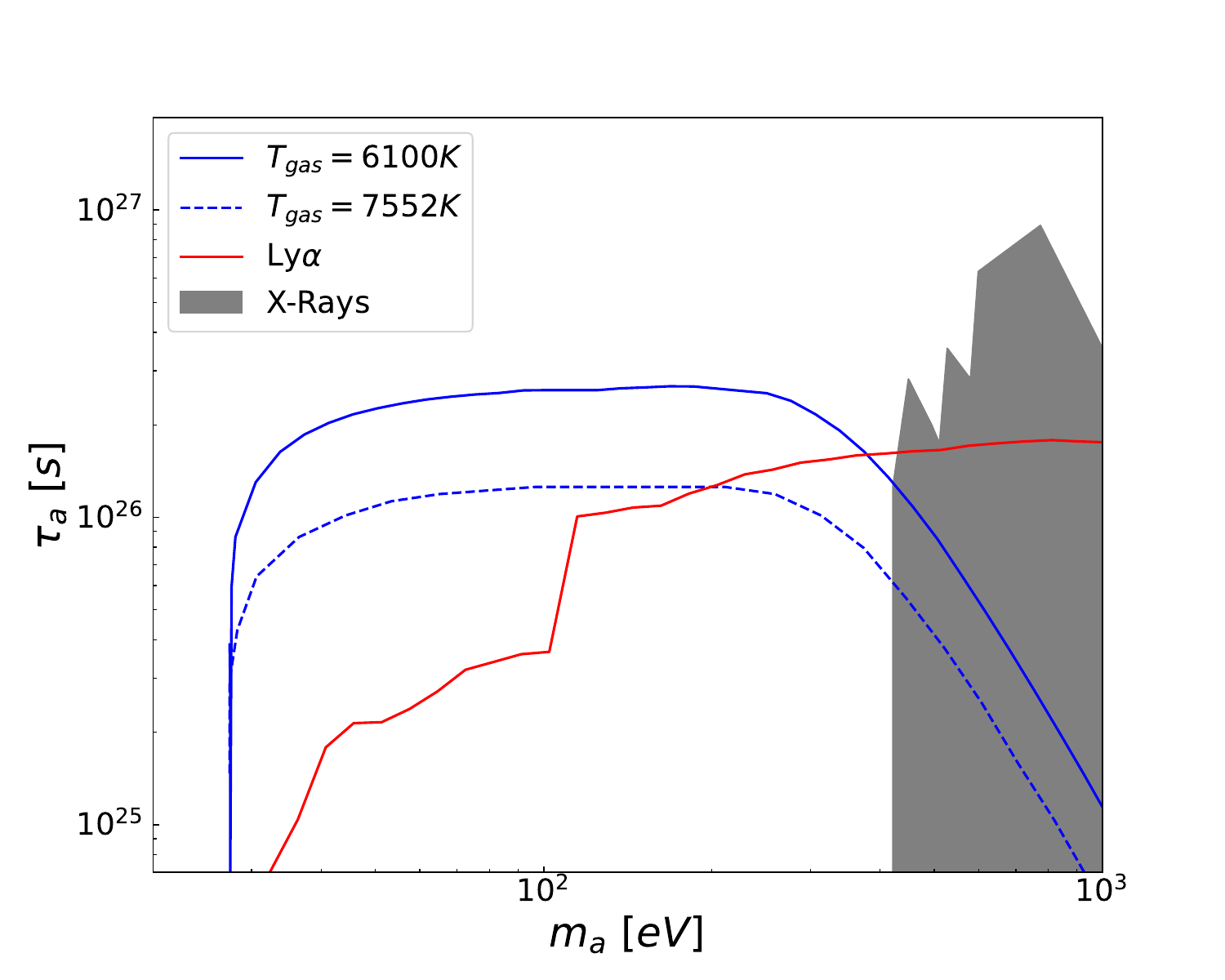}
\centering
\caption{Lyman-$\alpha$ limit on the plane of $(m_{a}, \tau_{a})$ at 95$\%$ CL,
where regions below the red curve are excluded.
For a comparison we have shown the other existing bounds including 
soft X-ray \cite{Cadamuro:2011fd} recasted from \cite{Boyarsky:2006fg, Boyarsky:2009ix} in shaded region, and Leo T \cite{Wadekar:2021qae} with two reference values of $T_{\rm{gas}}$ in blue. See text for details.}
\label{Lyc}
\end{figure}
%%%%%%%%%%%%%%%%%%%%%%%%%%%%%%%%%%%%%%%%%%%%%%%

$\underline{\mathbf{Comparison~with~existing~bounds}}$.
Fig.\ref{Lyc} shows the new Lyman-$\alpha$ limit on the plane of $(m_{a}, \tau_{a})$, 
compared to the other existing bounds in the literature. 
Three comments are in order: 
\begin{itemize}
\item Compared to reported Lyman-$\alpha$ limits \cite{Liu:2020wqz,Xu:2024uas} across the mass range of $m_{a}\geq$ keV,
we have extended it  (in red) down to the mass range of $m_{a}\sim 30$ eV in fig.\ref{Lyc}.
Note, this limit is absent in smaller $m_a$ region, as the DM decay induced photon energy $E_{\gamma}$ has to excess the threshold value of $\sim 10.2$ eV to heat the IGM.
\item For $m_{a}\sim 30-375$ eV, the Lyman-$\alpha$ limit is comparable with the Leo T bound \cite{Wadekar:2021qae} (in green) depending on the temperature of gas in Leo T, where we have used two reference values of $T_{\rm{gas}}=\{6100, 7552\}$ K for illustration. 
\item For $m_{a}\sim 375-425$ eV,  our result is the most stringent. 
\item For $m_{a}\sim 425-1000$ eV, this limit is complementary to soft X-ray bound \cite{Cadamuro:2011fd}  (in shaded region) derived from \cite{Boyarsky:2006fg, Boyarsky:2009ix}.
\end{itemize}
To summarize, the Lyman-$\alpha$ limit is the most stringent  in the ALP mass range of $m_{a}\sim 375-425$ eV and complementary to the Leo T and soft X-ray bound in the ALP mass range otherwise.

\section{Implications to ALP dark matter}
\label{imp}
Despite being insensitive to QCD axion DM whose surviving parameter region lies around $m_{a}<<1$ eV,  
the Leo T, soft-X ray and Lyman-$\alpha$ limit in fig.\ref{Lyc} have important physical implications to the ALP DM with mass $m_{a}\sim 30-1000$ eV under consideration. \\

$\underline{\mathbf{Misalignment~mechanism}}$. 
Under the context of misalignment mechanism \cite{Abbott:1982af, Preskill:1982cy,Dine:1982ah,Turner:1983he},
the ALP relic abundance is given by \cite{Arias:2012az}
\begin{eqnarray}\label{relicms}
\Omega_{a}h^{2}\sim 0.1 \left(\frac{m_{a}}{\rm{eV}}\right)^{1/2} \left(\frac{f_{a}}{10^{11}\rm{GeV}}\right)^{2} \left(\frac{m_{a}}{m_{a}(t_{*})}\right)^{1/2},
\end{eqnarray}
where $g_{a\gamma\gamma}=(\alpha/2\pi)f^{-1}_{a}$ with $f_{a}$ the broken scale of global $U(1)$ symmetry, 
\begin{eqnarray}\label{beta}
\frac{m_{a}}{m_{a}(t_{*})}=\left(\frac{\Lambda}{T(t_{*})}\right)^{\beta},
\end{eqnarray}
with $m_{a}(t)$ the effective ALP mass depending on time $t$, 
$t_{*}$ refers to the moment when $m_{a}(t_{*})=3H$ with $H$ the Hubble rate, 
$\Lambda\approx m_{a}f_{a}$ is the characteristic mass scale of unbroken $SU(N)$ gauge group, 
$T$ is the temperature of this new sector,
and $\beta$ is a dimensionless parameter. 
Eq.(\ref{relicms}) suggests that the observed DM relic density can be addressed in the parameter regions with $m_{a}\sim 1-10^{3}$ eV and $g_{a\gamma\gamma}\sim 10^{-15}-10^{-11}$ GeV$^{-1}$ depending on the value of $\beta$.

%%%%%%%%%%%%%%%%%%%%%%%%%%%%%%%%%%%%%%%%%%%%%%%%
\begin{figure}[htb!]
\centering
\includegraphics[width=15cm,height=15cm]{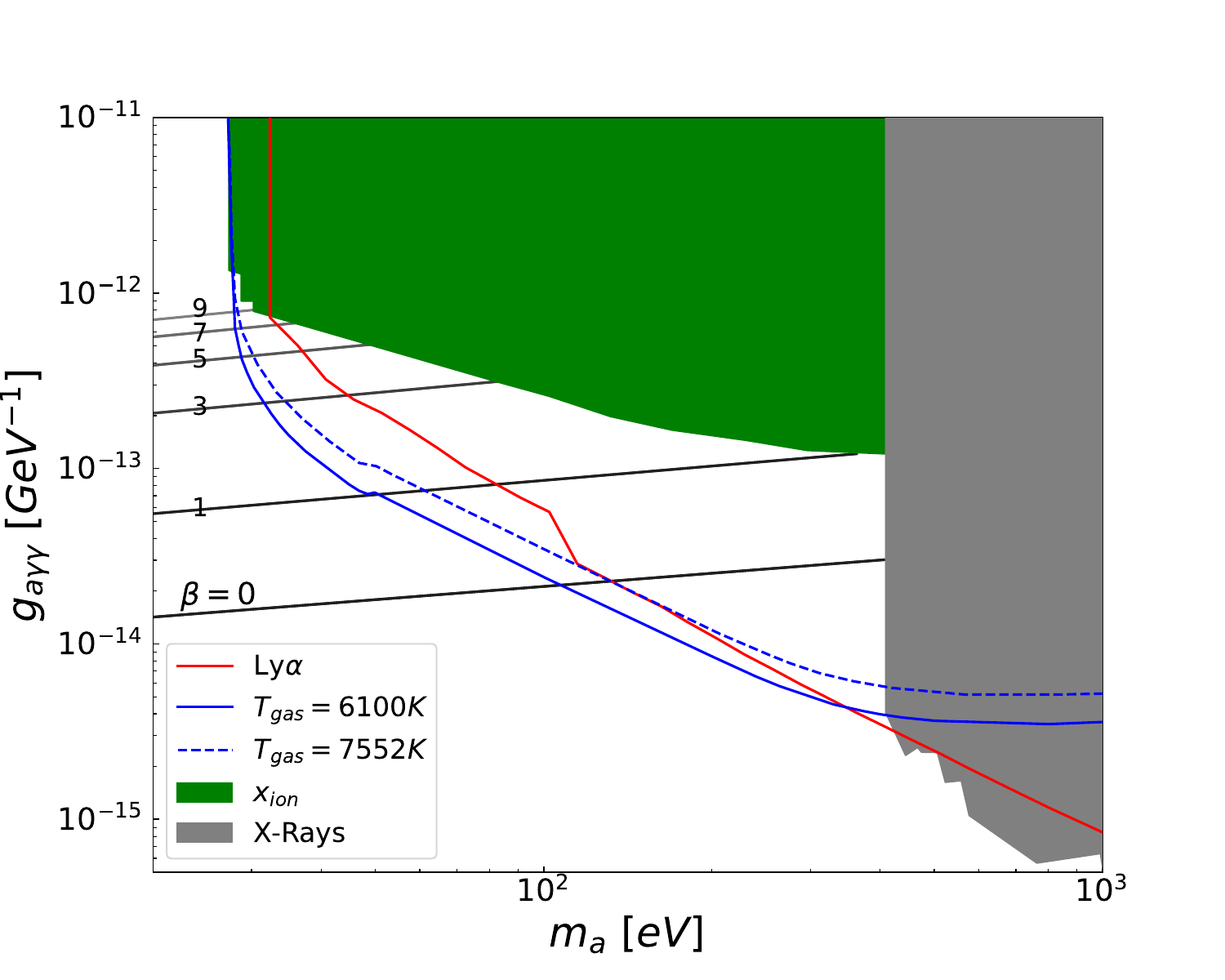}
\centering
\caption{Constraint on the ALP DM through the misalignment mechanism, compared to current experimental bounds from 
$x_{ion}$ \cite{Cadamuro:2011fd} (in green), soft X-ray \cite{Cadamuro:2011fd} (in gray), Leo T \cite{Wadekar:2021qae} (in blue), and Lyman-$\alpha$ forest (in red). 
The black lines correspond to the values of $\beta=\{0,1,3,5,7,9\}$ from bottom to top.
For a complete list of limits, see \cite{C. O’Hare}.}
\label{mis}
\end{figure}
%%%%%%%%%%%%%%%%%%%%%%%%%%%%%%%%%%%%%%%%%%%%%%%

Fig.\ref{mis} presents the constraints on the ALP DM from the misalignment mechanism, 
where the new Lyman-$\alpha$ limit in fig.\ref{Lyc} is converted to the plane of $(m_{a}, g_{a\gamma\gamma})$ using  eq.(\ref{lifetime}),
and the black lines refer to the values of $\beta=\{0,1,3,5,7,9\}$ from bottom to top.
Compared to the earlier exclusion results $m_{a}\leq \{400, 380, 90\}$ eV \cite{Arias:2012az} with respect to $\beta=\{0,1,3\}$ respectively made by ionization of primordial hydrogen ($x_{ion}$ in green) and  soft X-ray \cite{Cadamuro:2011fd} (in gray) bound,
they are updated to $m_{a}\leq \{110-140, 50-80, 30-45\}$ eV with respect to $\beta=\{0,1,3\}$ respectively in light of Leo T (in blue) and Lyman-$\alpha$ (in red) limit. 
These values can be can further improved by near future data of Lyman-$\alpha$ forest and 21-cm signal.\\

$\underline{\mathbf{Freeze-in~mechanism}}$. Apart from the misalignment mechanism, 
the ALP DM can be also produced via the freeze-in mechanism \cite{Hall:2009bx}. 
Given $g_{a\gamma\gamma}$ coupling,  the ALP relic abundance arises from Primakoff process $q \gamma\rightarrow qa$ where $q$ the SM charged fermions and inverse decay of $a\rightarrow \gamma\gamma$. 
The former is ultraviolet dominated being sensitive to reheating temperature $T_{\rm{reh}}$,
whereas the later is infrared dominated being independent on $T_{\rm{reh}}$. 
For $m_{a}<<$ MeV and $T_{\rm{reh}}>>10$ MeV considered here,
the former dominates over the later \cite{Jain:2024dtw, Langhoff:2022bij}.

%%%%%%%%%%%%%%%%%%%%%%%%%%%%%%%%%%%%%%%%%%%%%%%%
\begin{figure}[htb!]
\centering
\includegraphics[width=15cm,height=15cm]{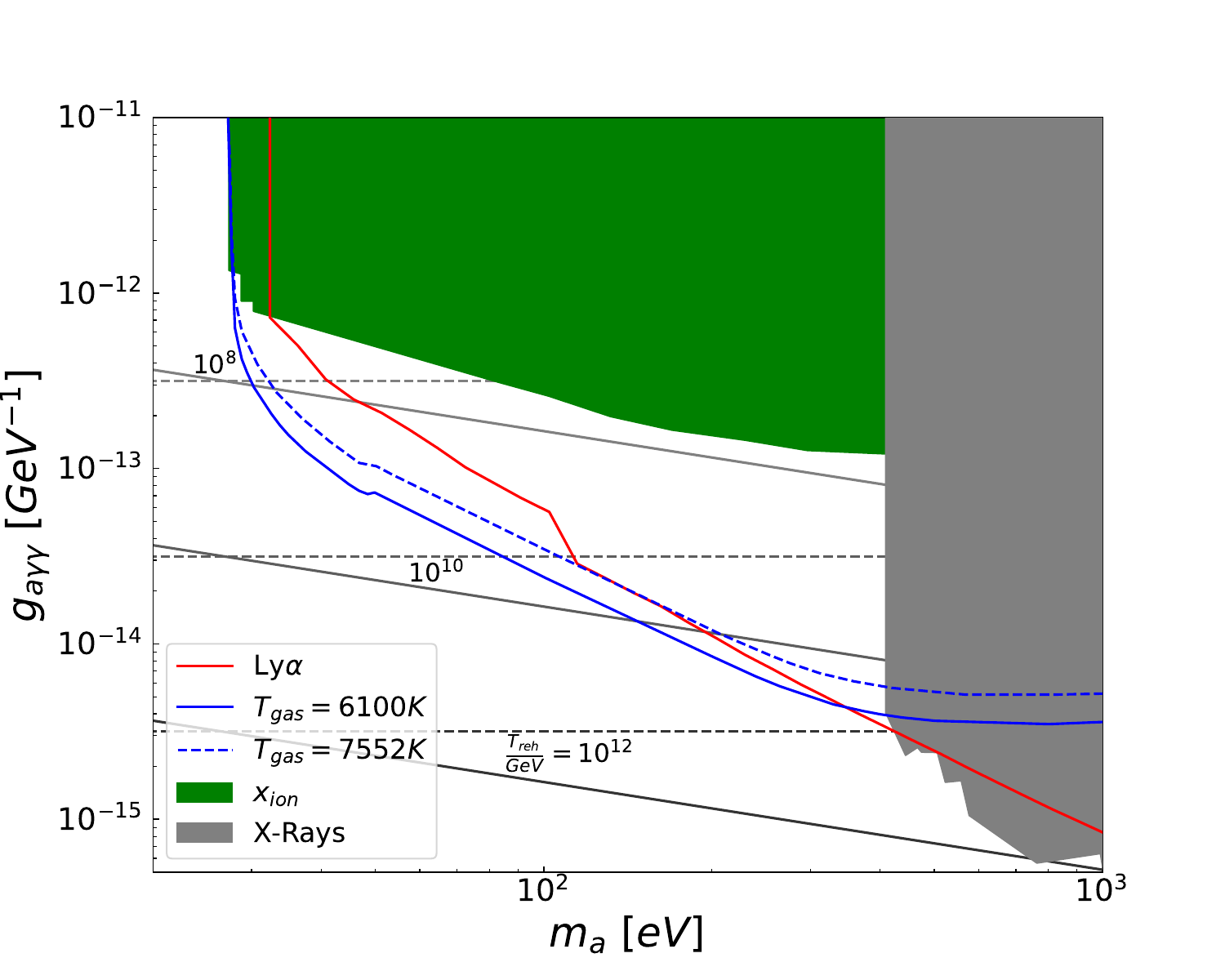}
\centering
\caption{Same as in fig.\ref{mis} but for the freeze-in mechanism. 
The solid gray lines refer to the observed DM relic abundance for various values of $T_{\rm{reh}}=\{10^{8}, 10^{10}, 10^{12}\}$ GeV from top to bottom. 
The dotted gray lines refer to the freeze-in condition in eq.(\ref{Tup}) with different values of $T_{\rm{reh}}$, 
above which the solid lines are excluded. }
\label{fi}
\end{figure}
%%%%%%%%%%%%%%%%%%%%%%%%%%%%%%%%%%%%%%%%%%%%%%%

Neglecting the inverse decay contribution, the ALP relic abundance due to the Primakoff process is given by 
\begin{eqnarray}
\label{rb}
\Omega_{a}h^{2}\approx 4.5 \times 10^{-5}\left(\frac{g_{a\gamma\gamma}}{10^{-11}~\rm{GeV}^{-1}}\right)^{2}
\left(\frac{m_{a}}{0.1~\rm{MeV}}\right)\left(\frac{T_{\rm{reh}}}{10~\rm{MeV}}\right) 
\end{eqnarray}
by scaling the result of \cite{Jain:2024dtw}.
The reheating temperature in eq.(\ref{rb}) is upper bounded as  \cite{Cadamuro:2011fd}
\begin{eqnarray}
\label{Tup}
T_{\rm{reh}}<T_{\rm{fo}}\approx 10^{5}\left(\frac{g_{a\gamma\gamma}}{10^{-11}~\rm{GeV}^{-1}}\right)^{-2} \rm{GeV},
\end{eqnarray}
where $T_{\rm{fo}}$ is the freeze-out temperature.
Eq.(\ref{rb}) shows that the observed value of DM relic density can be addressed by $g_{a\gamma\gamma}\sim 10^{-15}-10^{-11}$ GeV$^{-1}$ and $m_{a}\sim 1-1000$ eV depending on the value of $T_{\rm{reh}}$.

Fig.\ref{fi} shows the observed DM relic abundance (in solid gray) projected to the plane of $m_{a}-g_{a\gamma\gamma}$ for various values of $T_{\rm{reh}}=\{10^{8}, 10^{10}, 10^{12}\}$ GeV, which are compared to current experimental bounds and the freeze-in condition in eq.(\ref{Tup}) (in dotted gray).
With respect to each $T_{\rm{reh}}$, the solid gray line above the dotted gray line is excluded. 
For $T_{\rm{reh}}$ lower than $10^{12}$ GeV, the combination of Leo T and Lyman-$\alpha$ limit implies $m_{a}\leq \{30-35, 155-205\}$ eV for $T_{\rm{reh}}=\{10^{8}, 10^{10}\}$ GeV, respectively. 
For $T_{\rm{reh}}$ higher than $\sim 10^{12}$ GeV, 
the entire mass range considered is beyond the reaches of all current limits.
To be clarify, we remind the reader that ALP DM mass regions with $m_{a}< 1$ eV can be additionally excluded by limits not displayed in fig.\ref{fi},
a complete list of which can be found in ref.\cite{C. O’Hare}.

\section{Conclusion}
\label{con}
In this work we have presented the new Lyman-$\alpha$ limit within the DM mass range of $m_{a}\sim 30-1000$ eV.
It has been shown to be the most stringent in the ALP DM mass range of $m_{a}\sim 375-425$ eV and complementary in the otherwise ALP DM mass range, compared to the Leo T and soft-X ray limit.
Using these limits, we have excluded (i) the ALP DM mass down to $m_{a}/\rm{eV}\sim 110-140, 50-80, 30-45 $ eV for the parameter $\beta=0,1,3$ respectively in the context of misalignment mechanism, and (ii) the ALP DM mass down to $m_{a}/\rm{eV}\sim 30-35, 155-205$ for the reheating temperature $T_{\rm{reh}}/\rm{GeV}=10^{8}, 10^{10}$ respectively in the context of freeze-in mechanism.

\end{document}